\documentclass{ws-ijbc}
\begin{document}
\catchline{}{}{}{}{} 

\markboth{Bla{\v z} Krese et al.}{Experimental observation of a chaos-to-chaos transition in laser droplet generation}

\title{EXPERIMENTAL OBSERVATION OF A CHAOS-TO-CHAOS TRANSITION IN LASER DROPLET GENERATION}

\author{BLA{\v Z} KRESE,$^{\star}$ MATJA{\v Z} PERC,$^{\dagger}$ and EDVARD GOVEKAR$^{\star}$}
\address{$^{\star}$Laboratory of Synergetics, Faculty of Mechanical Engineering, University of Ljubljana\\
A{\v s}ker{\v c}eva cesta 6, SI-1000 Ljubljana, Slovenia\\
$^{\dagger}$Department of Physics, Faculty of Natural Sciences and Mathematics, University of Maribor\\
Koro{\v s}ka cesta 160, SI-2000 Maribor, Slovenia}

\maketitle

\begin{history}
\received{(to be inserted by publisher)}
\end{history}

\begin{abstract}
We examine the dynamics of laser droplet generation in dependence on the detachment pulse power. In the absence of the detachment pulse, undulating pendant droplets are formed at the end of a properly fed metal wire due to the impact of the primary laser pulse that induces melting. Eventually, these droplets detach, \textit{i.e.} overcome the surface tension, because of their increasing mass. We show that this spontaneous dripping is deterministically chaotic by means of a positive largest Lyapunov exponent and a negative divergence. In the presence of the detachment pulse, however, the generation of droplets is fastened depending on the pulse power. At high powers, the spontaneity of dripping is completely overshadowed by the impact of the detachment pulse. Still, amplitude chaos can be detected, which similarly as the spontaneous dripping, is characterized by a positive largest Lyapunov exponent and a negative divergence, thus indicating that the observed dynamics is deterministically chaotic with an attractor as solution in the phase space. In the intermediate regime, \textit{i.e.} for low and medium detachment pulse powers, the two chaotic states compete for supremacy, yielding an intermittent period-doubling to amplitude chaos transition, which we characterize by means of recurrence plots and their properties. Altogether, the transition from spontaneous to triggered laser droplet generation is characterized by a chaos-to-chaos transition with an intermediate dynamically nonstationary phase in-between. Since metal droplets can be used in various industrial applications, we hope that the accurate determination of the dynamical properties underlying their formation will facilitate their use and guide future attempts at mathematical modeling.
\end{abstract}


\section{Introduction}

\noindent The theory of dynamical systems and deterministic chaos \cite{schuster, strogatz, eckmann_ruelle} provides the backbone for our understanding of many natural and technological phenomena. The importance of nonlinearities inherent to many of them and the resulting ubiquitousness of deterministic chaos have led scientists and engineers of various fields to develop and use methods of nonlinear dynamics on observed data. Although sometimes still unnoticed, chaotic behavior \cite{gan}, fractal structures \cite{aguirreXrmp} and synchronization \cite{snychro, wang1, wang2} are deeply rooted in several fields of science \cite{abarbanel93, schreiber}. In contrast to the universal avowal of the chaos theory, the complexity and richness of the dynamics underlaying many technical processes often remains unexplored or is even completely overlooked. A vast potential in merging chaos theory with real life engineering systems lies within chaos control \cite{ottXprl, grchenXbook, ma1, ma2} to directly interfere with the system states. Due to the complexity of natural and technological processes, mathematical models are often nonexistent, so we face the problem of characterizing the process by means of analyzing experimental data. Nonlinear time series analysis \cite{abarbanel, kantz_schreiber} offers methods to determine dynamical properties of a particular system by analyzing the time series of one characteristic variable of the process. While these methods enable us to bridge the gap between the observed behavior and the theory of dynamical systems, we emphasize that the time series under study needs to meet conditions of having properties that are typical of deterministic systems \cite{kaplanXprl, kantz_schreiber}. Indeed, we point out the importance of verification whether the observed irregular behavior is deterministic and stationary in order to make the results of the nonlinear analysis meaningful.

In this paper we propose a set of experiments in order to study the dynamics of laser droplet generation, in particular the influence of the detachment pulse to it. Similarly to the traditional dripping faucet experiment \cite{faucet}, the surface tension and gravity force play a crucial role with the laser droplet generation. However the latter is governed by additional physical phenomena, including light-metal interaction, heating and phase transitions, which distinguish the two processes significantly. A laser pulse is used to melt the tip of the vertically placed metal wire. From the molten end a pendant droplet is formed due to the interplay between surface tension and gravity force. The droplet detaches when a surface tension force is overcome. This can either happen as a result of the droplet mass growth, or by means of intensive laser heating, which we apply by means of an additional detachment pulse at the end of the pendant droplet formation pulse. Here we present the significant influence of the detachment pulse power on the generation of laser droplets, in particular from the dynamics point of view. In order to do so we recorded a set of droplet generation sequences with various detachment pulse powers. The most important variable to observe during the process is the temperature of the wire end and the pendant droplet. We measure this indirectly by means of high-speed infrared (IR) camera. The time course of the temperature is finally obtained as the mean value over the pixel intensity of the IR snapshots. From a set of time series we selected three characteristic ones for the analysis. We start the analysis with the power spectra inspection and then continue with the nonlinear time series analysis. Applying the embedding theorem \cite{embedding1, embedding2} to reconstruct the phase space from a single variable, we use the mutual information \cite{fraseXpra} and the false nearest neighbor \cite{kennelXpra} methods to obtain optimal embedding parameters. A determinism test \cite{kaplanXprl} follows as well as testing for nonstationarity using recurrence plots and their quantification \cite{recurplots1, recurplots2}. At the end we calculate the spectra of Lyapunov exponents \cite{briggsXpla, lyapspec2} for the time series which exhibited deterministic and stationary properties. We observe a chaos-to-chaos transition with an intermediate dynamically nonstationary phase in-between as the detachment pulse power is increased. Finally, we outline the significance of our analysis for the deeper understanding of the process itself, as well as for future attempts at mathematical modeling.

The paper is structured as follows. In Section $2$ an accurate description of the experimental setup and experiments is given. Section $3$ is devoted to presenting the results of nonlinear time series analysis, while in the last section we summarize our findings and conclude the paper.
                  	
\section{Experimental Setup}

\noindent We use a laser pulse as a source of energy in order to generate droplets from the metal wire. The process phenomenologically consists of two phases, \textit{i.e.} the generation of the pendant droplet and its detachment. In the first phase a primary pulse is used to melt the end of a vertically fed metal wire. From the molten end a pendant droplet is formed due to the action of surface tension and gravity force. The surface tension drags the pendant droplet up the wire so the wire has to be properly fed to obtain a certain droplet volume and to ensure a proper relative position of the laser beam with respect to the wire. Now a droplet, undulating at the tip of the wire, needs to detach. To achieve the detachment, the surface tension force has to be overcome. One way to reach the threshold of detachment is by droplet mass growth, but in our case we attached an additional secondary pulse, \textit{i.e.} the detachment pulse, at the end of the pendant droplet formation phase in order to stimulate the detachment of the pendant droplet. Notably, metal droplets are being used in many manufacturing applications, like for instance droplet joining, where a molten droplet is placed onto the joining spot \cite{join, join1, join2}. Other potential applications include the generation of 3D structures accomplished by selective deposition of droplets into layers and micro casting. Laser droplet generation is a process which encompasses the most vital characteristics needed for these technologies. To be able to effectively optimize and control the process it is essential to know its dynamics, which we aim to determine from experimental data.

\begin{figure}
\begin{center} \includegraphics[width=8.0cm]{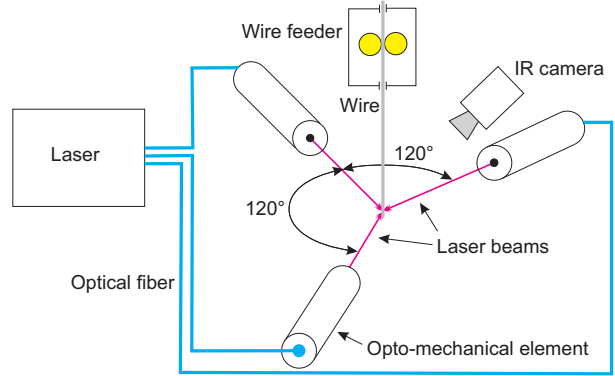}
\caption{\label{experiment} Schematic presentation of the experimental setup. The temperature is measured indirectly by means of a high-speed infrared (IR) camera (see main text for details).}
\end{center}
\end{figure}

For the purpose of studying laser droplet generation we have developed an experimental system that is schematically presented in Fig.~\ref{experiment}. The Nd:YAG pulse laser, opto-mechanical elements, the wire feeder and the infrared camera are the main parts of the experimental system. The Nd:YAG laser is used for generating laser pulses with a wavelength of $1.06$~$\mu$m. The maximal laser pulse power is $8$~kW and the pulse duration needs to be between $0.3$~ms and $20$~ms. The maximal pulse repetition rate is $300$~Hz with an average power of $0.25$~kW. The uniform heating of the wire and process symmetry (see Fig.~\ref{experiment}) are achieved by division of the laser light into three equal laser beams. By means of the opto-mechanical elements the beams are distributed equiangular along the wire circumference and perpendicularly focused onto the wire's surface. The wire is fed vertically by means of a controlled wire feeder having a triangular velocity profile, which does not vary with the detachment dynamics and is applied in order to synchronize the triggering of the laser pulse with the stepwise wire feeding. Since the temperature is the most important variable of the process it was indirectly measured by means of a high-speed infrared camera. Given the properties of the light emitted by the wire end and the pendant droplets, the snapshots were acquired at wavelengths between $3.5$~$\mu$m and $5$~$\mu$m.

\begin{figure}
\begin{center} \includegraphics[scale=1]{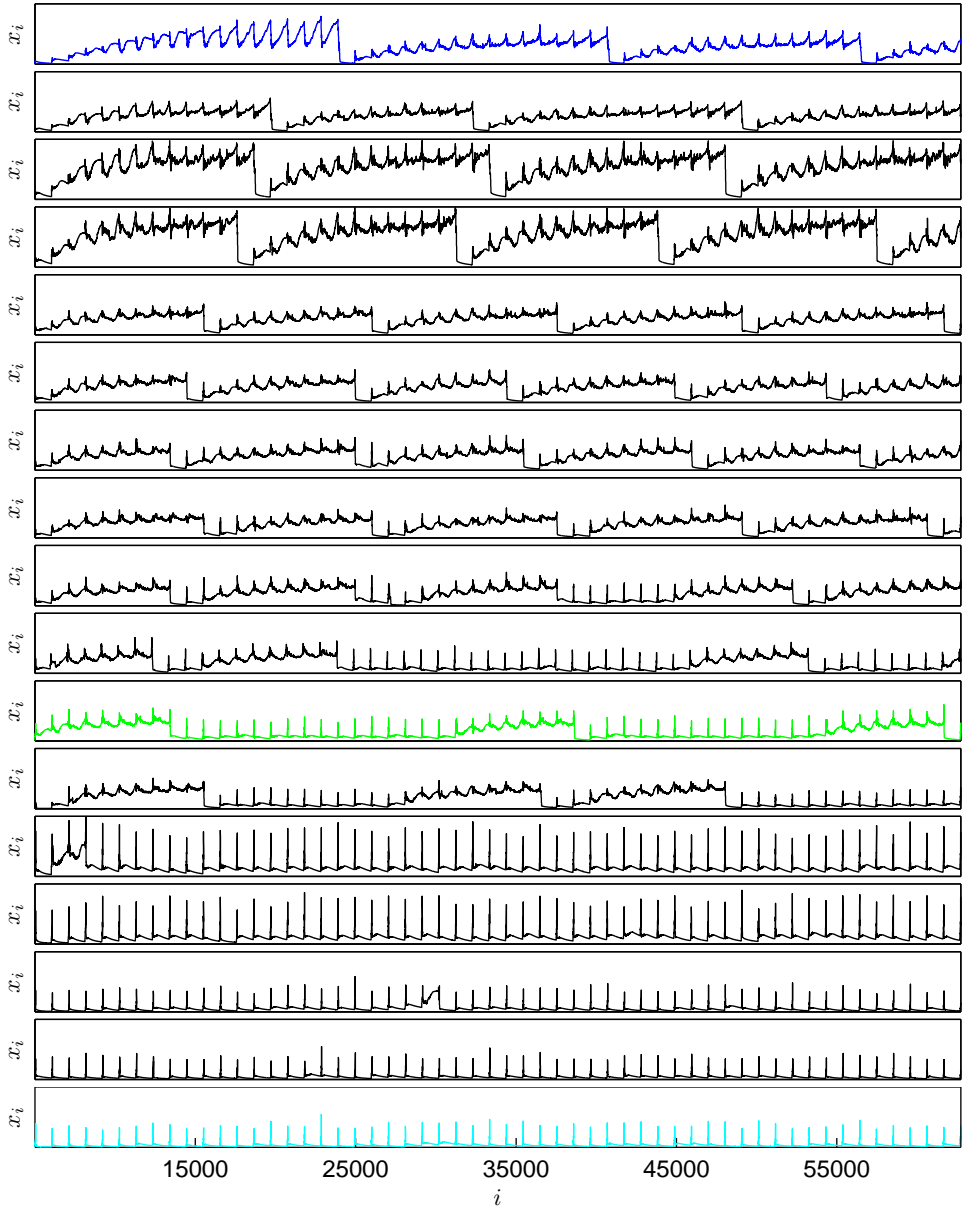}
\caption{\label{time} Characteristic excerpts of time courses, capturing the process of laser droplet generation via the pixel intensity of high-speed infrared snapshots, for different powers of the detachment pulse. From top to bottom the detachment pulse power $P_{dp}$ was increased from 0~kW to 8~kW via increments of 0.5~kW. In what follows, we will focus on the time courses obtained for 0~kW (blue), 5~kW (green) and 8~kW (cyan). Prior to the analysis the three time courses were subject to Wiener filtering, removing the high-frequency noisy component that is due to the infrared imaging, and were rescaled to the unit interval for simplicity. Note also that the depicted traces were recorded at twice the sampling frequency that was subsequently used for the time series analysis presented in Section 3.}
\end{center}
\end{figure}

According to the given description of the experimental setup there are several parameters that influence the process of laser droplet generation. Here we present those that were important for carrying out the experiments. We used a nickel wire of diameter $0.6$~mm. A rectangular laser pulse of power $1.44$~kW and duration of $12$~ms was used as a primary pulse in order to form a pendant droplet. Subsequently a detachment pulse of $1.2$~ms duration and various powers was attached to the primary pulse with a delay of $2.0$~ms. The power of the detachment pulse was varied from $0$~kW to $8$~kW with a step of $0.5$~kW. The so composed laser pulse was being triggered with a frequency of $3$~Hz. The sampling frequency of the infrared camera was $1428$~Hz (except for the time series presented in Fig.~\ref{time}, where we have used $3147$~Hz) at a snapshot size of $32 \times 64$ pixels. Finally, the spatiotemporal temperature field was converted into a single scalar time series by calculating the mean value of the pixel intensity of every snapshot. A set of resulting time series in dependence on the power of the detachment pulse is shown in Fig.~\ref{time}.

\section{Time Series Analysis}

\begin{figure}
\begin{center} \includegraphics[scale=1]{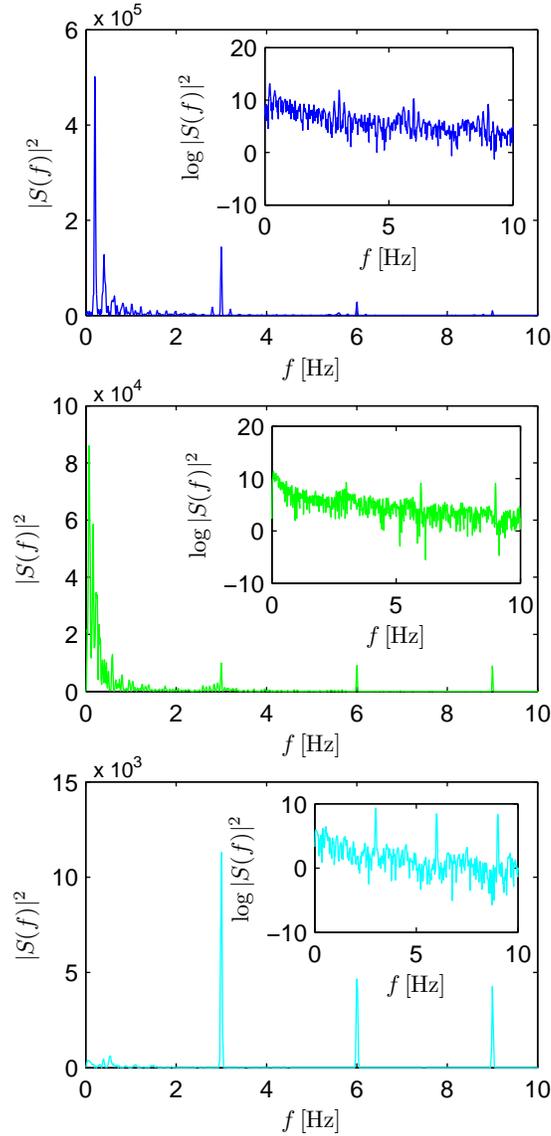}
\caption{\label{power} Power spectra of the three examined time courses. From top to bottom the detachment pulse powers are $P_{dp}=0, 5$ and $8$~kW, respectively. Insets feature the power spectra in logarithmic scale. Notice the continuity in all the spectra, visible especially good on logarithmic scale, albeit the periodic impact of detachment pulses becomes clearly visible in the bottom panel (harmonic spikes). Still, the power spectra hint toward deterministically chaotic behavior, as we will show using nonlinear methods of time series analysis in what follows.}
\end{center}
\end{figure}

\noindent We begin with visual inspection of acquired time courses of the temperature profile, as obtained from the high-speed infrared images, for different detachment pulse powers in Fig.~\ref{time}. It can be observed that the dynamics changes rather dramatically from the top ($P_{dp}=0$~kW) to the bottom panel ($P_{dp}=8$~kW). In the upper-most series, low and high frequency components can be inferred, which can be linked nicely with the two-phase process of spontaneous laser droplet generation. Namely, the high frequency oscillations correspond to droplet volume (mass) and temperature growth, which is followed by a sudden drop of the signal amplitude due to the spontaneous droplet detachment, giving rise to the low frequency component. On the other hand, in the lower-most panel a single frequency dominates, which is that of the detachment pulse triggering. Note that for high detachment pulse powers the droplet detaches virtually every time the laser pulse is triggered, thus completely overriding the spontaneous growth of droplet volume and mass that can be observed in the upper-most series (high-frequency small-amplitude undulations). For intermediate detachment pulse powers, however, droplets detached only occasionally following the triggering of the laser pulse, while sometimes they remain undulating and acquiring mass via the spontaneous dripping mechanism. The result is a mixture between spontaneous and forced dripping, manifesting as what appears to be an intermittent dynamical state between two complex behaviors (see \textit{e.g.} the time course colored green in Fig.~\ref{time} depicting the dynamics recorder at $P_{dp}=5$~kW). In what follows, we will use methods of nonlinear time series analysis in order to quantify the dynamics for different detachment pulse powers, focusing specifically on time courses obtained for $P_{dp}=0$~kW (blue), $P_{dp}=5$~kW (green) and $P_{dp}=8$~kW (cyan). This coloring for the three considered series will be used throughout this work.

Before commencing with nonlinear time series analysis, however, it is instructive to have a look at the power spectra of the three series to get an impression about their complexity. Figure~\ref{power} features the obtained results in linear and logarithmic scale. Especially in logarithmic scale the continuity of spectra in all three cases is visible very well, thus suggesting that the behavior is, besides being characterized by some predominant frequencies, inherently irregular. Harmonic spikes dominate for $P_{dp}=8$~kW, thus indicating a strong periodic component, which is due to the periodic action of the detachment pulse. In-between the two extreme cases the power spectrum is a mixture of both, on one hand having a somewhat sharper periodic component than the $P_{dp}=0$~kW case, but on the other still having significantly more continuity and a much stronger low-frequency component (due to the occasional spontaneous dripping) than the $P_{dp}=8$~kW case. Results presented in Fig.~\ref{power} thus support our visual assessment of the dynamics based on time courses in Fig.~\ref{time}, but also clearly outline the necessity for more sophisticated methods of analysis based on nonlinear statistics.

\begin{figure}
\begin{center} \includegraphics[scale=1]{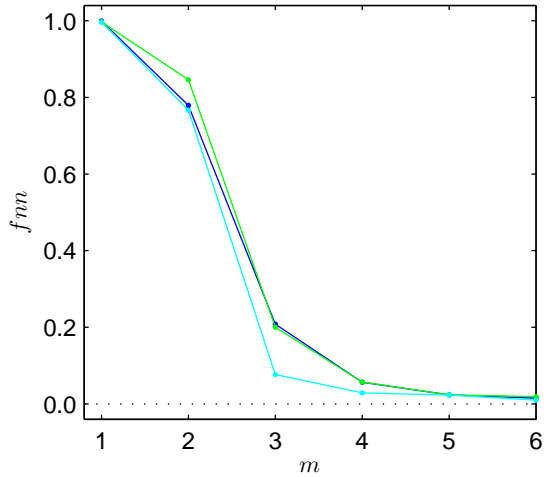}
\caption{\label{fnn} Determination of the minimally required embedding dimension. The fraction of false nearest neighbors ($fnn$) drops close ($<0.01$) to zero at $m = 5$ for all three time series. For the determination of false nearest neighbors \cite{kennelXpra}, we have used the first minimum of the mutual information for the embedding delay $\tau$ \cite{fraseXpra}. Specifically, the values were $\tau=260, 21$ and $13$ for detachment pulse powers $P_{dp}=0, 5$ and $8$~kW, respectively.}
\end{center}
\end{figure}

\begin{figure}
\begin{center} \includegraphics[scale=1]{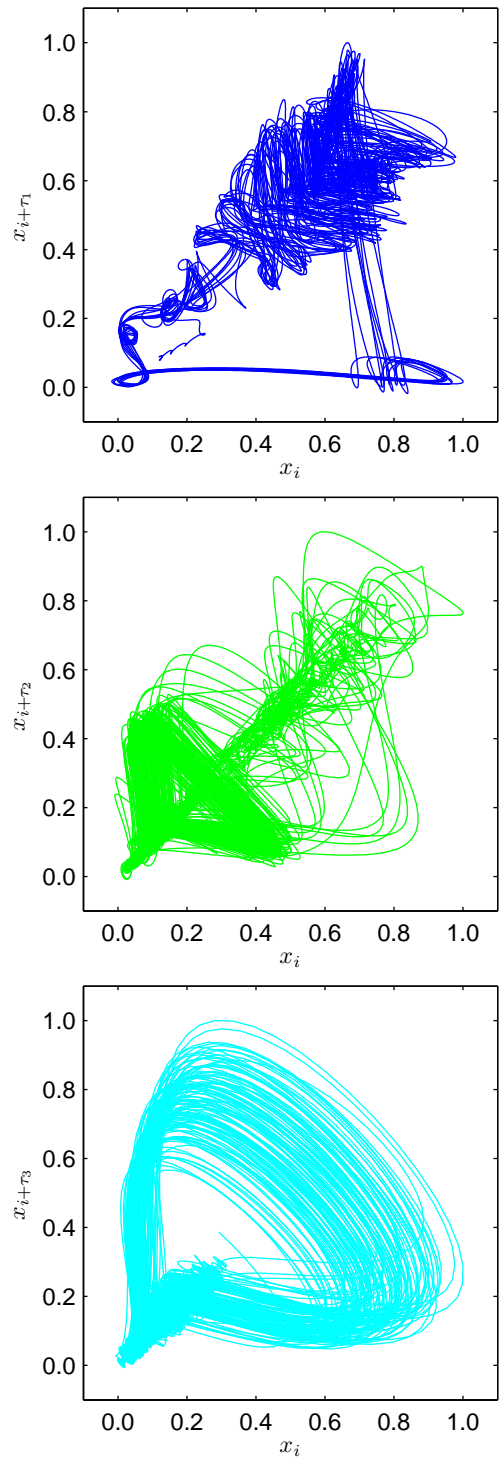}
\caption{\label{phase} Two-dimensional projections of the phase space for $P_{dp}=0$~kW (top panel), $P_{dp}=5$~kW (middle panel) and $P_{dp}=8$~kW (bottom panel). Reconstruction parameters were those stated in the caption of Fig.~\ref{fnn}. Determinism factor of all three phase spaces, determined according to the algorithm proposed by Kaplan and Glass \cite{kaplanXprl}, is $\kappa >0.9$, thus confirming the deterministic nature of the examined laser droplet generation dynamics.}
\end{center}
\end{figure}

Underlying almost all methods of nonlinear time series analysis is the embedding theorem \cite{embedding1, embedding2}, which states that for a large enough embedding dimension $m$ the delay vectors
\begin{equation}
{\rm \bf z}(i)=[x_i, x_{i+\tau}, x_{i+2 \tau}, \dots, x_{i+(m-1)\tau}]
\label{embed}
\end{equation}
yield a phase space that has exactly the same properties as the one formed by the original variables of the system. In Eq.~(\ref{embed}) variables $x_i$, $x_{i+\tau}$, $x_{i+2 \tau}$,$\dots$, $x_{i+(m-1)\tau}$ denote values (rescaled to the unit interval for simplicity) of the indirectly measured temperature at times $t=i{\rm d}t$, $t=(i+\tau){\rm d}t$, $t=(i+2\tau){\rm d}t$,$\dots$, $t=[i+(m-1)\tau]{\rm d}t$, respectively, whereby $\tau$ is the embedding delay and ${\rm d}t$ is the sampling time of data points equaling $7 \cdot 10^{-4}$~s in all three cases. However, while the implementation of Eq.~(\ref{embed}) is straightforward, we first have to determine proper values for the embedding parameters $m$ and $\tau$. For this purpose, the mutual information \cite{fraseXpra} and the false nearest neighbor method \cite{kennelXpra} can be used, respectively. Since the mutual information between $x_i$ and $x_{i+\tau}$ quantifies the amount of information we have about the state $x_{i+\tau}$ presuming we know $x_i$ \cite{shaw}, Fraser and Swinney \cite{fraseXpra} proposed to use the first minimum of the mutual information as the optimal embedding delay. Results for the three considered detachment pulse powers are stated in the caption of Fig.~\ref{phase}. The false nearest neighbor method, on the other hand, relies on the assumption that the phase space of a deterministic system folds and unfolds smoothly with no sudden irregularities appearing in its structure. By exploiting this assumption one comes to the conclusion that points that are close in the reconstructed embedding space have to stay sufficiently close also during forward iteration. If a phase space point has a close neighbor that does not fulfil this criterion it is marked as having a false nearest neighbor. As soon as $m$ is chosen sufficiently large, the projection effects due to a mapping of the time series onto a space with too few degrees of freedom should disappear, and with them the fraction of points that have a false nearest neighbor ($fnn$) should converge to zero \cite{kennelXpra}. Note that the method implicitly assumes that a deterministic time series is given as input. This, however, cannot be taken for granted, and indeed a simple extension of the originally proposed false nearest neighbor method \cite{heggerXpre} can be used also as a determinism test. Here we employ the classical algorithm proposed by Kennel \textit{et al.} \cite{kennelXpra} and use the determinism test due to Kaplan and Glass \cite{kaplanXprl}. Results of the false nearest neighbor method are presented in Fig.~\ref{fnn}, showing that $fnn \rightarrow 0$ at $m=5$ for all three cases. We will thus use these values as input for Eq.~(\ref{embed}) in what follows.

Having all the parameters at hand for reconstructing the phase space from the observed variable (see Fig.~\ref{phase}), we can proceed by employing the determinism test proposed by Kaplan and Glass \cite{kaplanXprl}. The test is simple but effective, measuring average directional vectors in a coarse-grained embedding space. The idea is that neighboring trajectories in a small portion of the embedding space should all point in the same direction, thus assuring uniqueness of solutions in the phase space, which is the hallmark of determinism. To perform the test, the embedding space has to be coarse grained into equally sized boxes. The average directional vector pertaining to a particular box is then obtained as follows. Each pass $p$ of the trajectory through the $k$-th box generates a unit vector ${\rm \bf e}_{p}$, whose direction is determined by the phase space point where the trajectory first enters the box and the phase space point where the trajectory leaves the box. The average directional vector ${\rm \bf V}_{k}$ through the $k$-th box is then
\begin{equation}
{\rm \bf V}_{k}=n^{-1}\sum_{p=1}^n {\rm \bf e}_{p}
\label{vf}
\end{equation}
where $n$ is the number of all passes through the $k$-th box. Completing this task for all occupied boxes gives us a directional approximation for the vector field. If the time series originates from a deterministic system, and the coarse grained partitioning is fine enough, the obtained directional vector field ${\rm \bf V}_{k}$ should consist solely of vectors that have unit length. Hence, if the system is deterministic, the average length of all the directional vectors $\kappa$ will be close to one. The determinism factor pertaining to the five-dimensional embedding spaces presented in Fig.~\ref{phase} that were coarse grained into a $12 \times 12 \times \ldots \times 12$ grid is $\kappa>0.9$, which confirms the deterministic nature of all three studied time series.

By now we have successfully reconstructed the phase space from the observed time courses and established their deterministic origin. In the continuation it would be possible to apply methods of nonlinear time series analysis that yield invariant quantities of the system, such as for example Lyapunov exponents \cite{briggsXpla, bryantXprl, abarbanelXjns, lyapspec1, lyapspec2} or dimension estimates \cite{grassbergerXprl, theilerXpra, kantzXchaos}, in order to obtain deeper insights into the system dynamics. However, these quantities could be meaningless if the studied time courses did not originate from a stationary system. Thus, in order to justify further analysis, we have to verify if the studied series possess properties that are typical of stationary courses.

\begin{figure}
\begin{center} \includegraphics[scale=1]{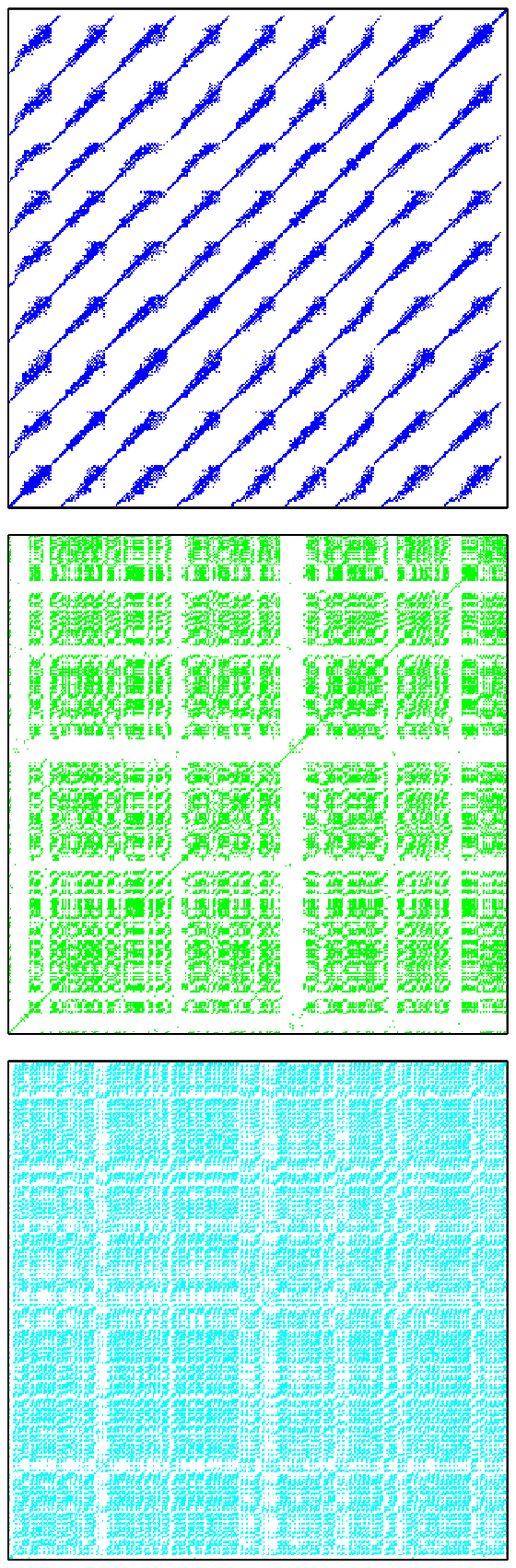}
\caption{\label{recur} Recurrence plots of the three examined time courses. From top to bottom the detachment pulse powers are $P_{dp}=0, 5$ and $8$~kW, respectively. For each time series we have selected $\epsilon$ such that the recurrence rate was approximately $1 \%$, which means $75 \%$, $15 \%$ and $30 \%$ of the standard deviation of the phase space from top to bottom, respectively. Note the obvious nonstationary dynamics in the middle panel, which is a consequence of the transition from spontaneous to triggered dripping. At $P_{dp}=5$~kW both processes play a noticeable role in the overall system dynamics, thus making its characterization via invariants, such as the Lyapunov exponents, questionable.}
\end{center}
\end{figure}

\begin{figure}
\begin{center} \includegraphics[scale=1]{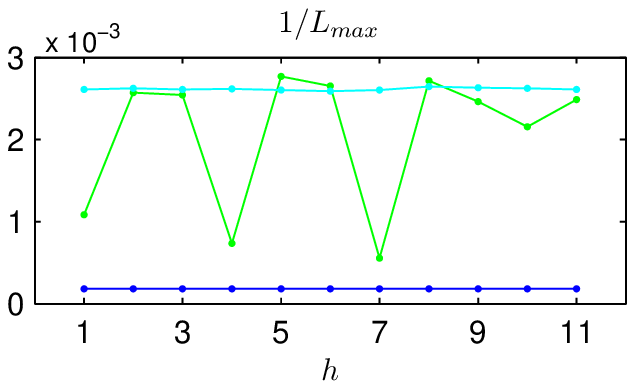}
\caption{\label{ana} The recurrence quantification analysis of recurrence plots presented in Fig.~\ref{recur}. Depicted are the inverse lengths $1/L_{max}$ of the longest diagonal line, whereby the three time courses were partitioned into 11 equally long non-overlapping segments each. While for $P_{dp}=0$~kW (blue) and $P_{dp}=8$~kW (cyan) the values vary insignificantly over the windows, for $P_{dp}=5$~kW (green) the nonstationarity is obvious. Indeed, $1/L_{max}$ jumps between the two extreme cases of spontaneous (blue) and triggered (cyan) dripping, thus evidencing nicely the transition taking place between the two dynamical states.}
\end{center}
\end{figure}

An appealing and simple graphical tool that enables the assessment of stationarity in an observed system is the recurrence plot \cite{recurplots1, recurplots2}. Recurrent behavior is an inherent property of oscillating systems. For regular oscillators time-distinct states in the phase space can be arbitrarily close, \textit{i.e.} $\|{\rm \bf z}(i)-{\rm \bf z}(j)\| = 0$ if times $i$ and $j$ differ exactly by some integer of the oscillation period, whereas for chaotic systems this distance is always finite. The recurrence plot is a two-dimensional square-grid graph with time units on both axes, whereby, in the most common case \cite{recurplots1, recurplots2}, points ($i,j$) that satisfy $\|{\rm \bf z}(i)-{\rm \bf z}(j)\|< \epsilon$ are marked with color while all others are left white. It is worth noting that depending on the application, there also exist several variations of recurrence plots that can be used for determining various properties of observed dynamics \cite{recurplots3, recurplots4, recurplots4_1, recurplots5, recurplots6, recurplots7, recurplots8, rpx1, rpx2}. For the visual assessment of recurrence plots the most important features are the large and small scale structure, latter being termed typology and texture \cite{recurplots1}, respectively. By visually inspecting the typology and texture of a recurrence plot, properties of the system such as stationarity and determinism can be assessed. In particular, a homogenous typology is an indicator that the studied data set originated from a stationary process. Contrary, a non-homogenous or disrupting typology indicates non-stationarity in the system. Texture, on the other, can provide information regarding deterministic vs. stochastic origin of the signal, as well as give insights on the complexity of oscillations. Lack of texture, \textit{i.e.} solely isolated recurrence points, often indicate stochastic origin of the examined time series, while diagonal lines indicate deterministic oscillations, which depending on the complexity of emerged small-scale patterns can be further classified into simple or chaotic oscillations. The recurrence plots of the three studied time courses are presented in Fig.~\ref{recur}. It can be observed that for $P_{dp}=0$~kW (top panel; blue) and $P_{dp}=8$~kW (bottom panel; cyan) the typology is homogenous, while for $P_{dp}=5$~kW (middle panel; green) it is not. In particular, several thick horizontal and vertical white lines disrupt the otherwise fairly homogenous squares lying in-between. From this it can be concluded that the recording for $P_{dp}=0$~kW and $P_{dp}=8$~kW stem from a dynamically stationary process, while the recording for $P_{dp}=5$~kW is most likely nonstationary. To strengthen this visual assessment, we have determined also the length of the longest diagonal $L_{max}$ in $11$ equally long segments in each of the three time courses. Results presented in Fig.~\ref{ana} clearly attest to the fact that, while for $P_{dp}=0$~kW (blue) and $P_{dp}=8$~kW (cyan) the dynamics is the same in all segments, for $P_{dp}=5$~kW (green) this is not the case as indeed $1/L_{max}$ jumps between the two extreme cases of spontaneous (blue) and triggered (cyan) dripping. From this we conclude that only the time courses obtained for $P_{dp}=0$~kW and $P_{dp}=8$~kW are both deterministic and dynamically stationary, while the one for $P_{dp}=5$~kW is deterministic but nonstationary. This in turn implies that the transition between spontaneous and forced (triggered by means of the detachment pulse) dripping is characterized by an intermittent mixture of the two extreme cases, whereby the forced dynamics is the more prevalent the higher the power of the detachment pulse.

\begin{figure}
\begin{center} \includegraphics[scale=1]{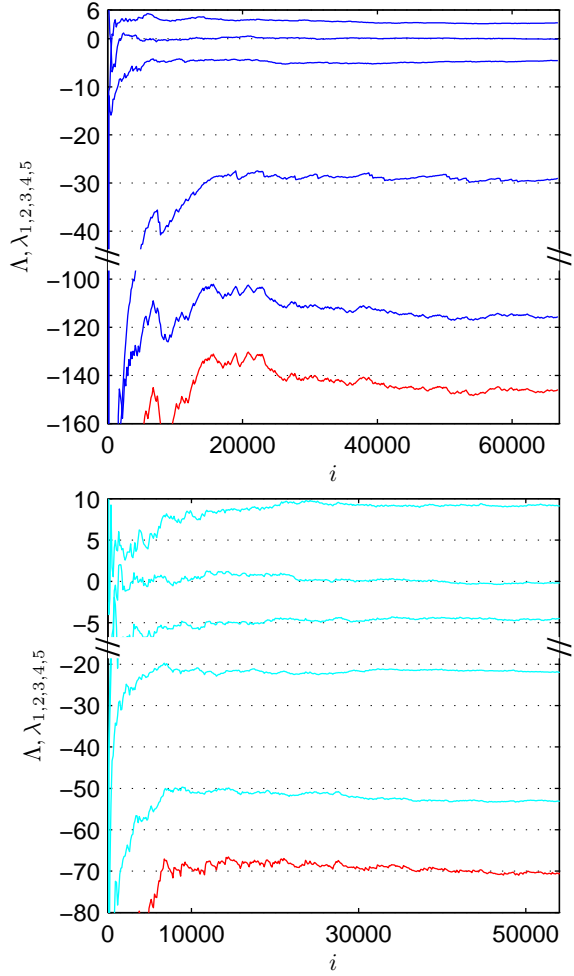}
\caption{\label{lyapspec} Spectra of Lyapunov exponents determined using radial basis functions for the approximation of the flow. Top panel shows results for $P_{dp}=0$~kW and the bottom panel depicts results for $P_{dp}=8$~kW. In both panels from top to bottom the lines depict the convergence of the largest ($\lambda_{1}$) to the smallest ($\lambda_{5}$; most negative) Lyapunov exponent as a function of the discrete time $i$. The lowest line (red) shows the sum of all five exponents, \textit{i.e.} the divergence $\Lambda=\sum_{j=1..m}\lambda_{j}$. A linear fit towards the end of the curves gives for the top panel $\lambda_{1} = (3.2 \pm 0.1){\rm s^{-1}}$, $\lambda_{2} = (0.0 \pm 0.1){\rm s^{-1}}$ and $\Lambda = -(145 \pm 3){\rm s^{-1}}$, while for the bottom panel we have $\lambda_{1} = (9.2 \pm 0.2){\rm s^{-1}}$, $\lambda_{2} = (0.0 \pm 0.2){\rm s^{-1}}$ and $\Lambda = -(70 \pm 2){\rm s^{-1}}$. Note that the vertical axis has a break in both panels.}
\end{center}
\end{figure}

Finally, it remains of interest to accurately quantify the dynamics of the two time courses that we found to be both deterministic as well as stationary. For this purpose we calculate the spectra of Lyapunov exponents $\lambda_j$ where $j=1, 2, \ldots, m$, knowing with reasonable certainty that the obtained results are due to deterministic nonlinear dynamics rather than noise or varying systems parameters during data acquisition. We employ radial basis functions for the approximation of the flow in the phase space. Using the phase space reconstruction parameters obtained above, $M=10$ nearest neighbors of each ${\rm \bf z}(i)$ to make the fit, and the stiffness parameter $r=7$ \cite{lyapspec2}, the exponents change their sign upon time reversal of the flow and converge robustly as the number of iterations increases. Figure~\ref{lyapspec} features the individual convergence curves, from which we obtain, for the top panel ($P_{dp}=0$~kW) $\lambda_{1} = (3.2 \pm 0.1){\rm s^{-1}}$, $\lambda_{2} = (0.0 \pm 0.1){\rm s^{-1}}$ and the divergence as the sum over all $\lambda_j$ equal to $\Lambda = -(145 \pm 3){\rm s^{-1}}$, while for the bottom panel ($P_{dp}=8$~kW) we have $\lambda_{1} = (9.2 \pm 0.2){\rm s^{-1}}$, $\lambda_{2} = (0.0 \pm 0.2){\rm s^{-1}}$ and $\Lambda = -(70 \pm 2){\rm s^{-1}}$. From the positive largest Lyapunov exponent, the vanishing second Lyapunov exponent, and the negative divergence, we can conclude that the dynamics of laser droplet generation, irrespective of whether it is spontaneous or forced by means of a strong detachment pulse, is deterministically chaotic, and that there exists a stable attractor in the phase space to which any given cloud of initial condition converges in time. A distinctive property of the two chaotic states is that the forced dynamical behavioral has a strong periodic component with noticeable amplitude modulation, \textit{i.e.} amplitude chaos, while the spontaneous dripping is primarily frequency modulated, \textit{i.e.} period-doubling chaos. The transition from spontaneous to the forced laser droplet generation is thus characterized by a chaos-to-chaos transition with an intermittent dynamically nonstationary phase in-between.

\section{Summary}

\noindent We have examined an experimental setup with the aim of determining the dynamics of laser droplet generation in dependence on the detachment pulse power. Using a high-speed infrared camera, we have indirectly measured the spatiotemporal profile of temperature of the molten end of the wire and the pending droplets. Subsequently, the time courses were obtained as the mean value over the pixel intensity of every infrared snapshot, and analyzed systematically with methods of linear and nonlinear time series analysis. After reconstructing the phase space from the observed variables, we have verified that the later have properties that are typical for deterministic systems. We have shown that the minimally required embedding dimension is five, which altogether suggests that it would be justified to mathematically model the process of laser droplet generation with no more than five first-order ordinary differential equations. Subsequently, we have constructed and quantified recurrence plots to show that only the fully spontaneous and fully forced time courses are dynamically stationary, while in the region of intermediate detachment pulse powers the dynamics is nonstationary. Accordingly, we have determined the whole spectra of Lyapunov exponents for the two extreme cases by approximating the flow in the phase space with radial basis functions. Our calculations revealed that the largest Lyapunov exponent is positive, the second is zero, while the divergence is negative, thus obtaining strong indicators that the observed dynamics, either spontaneous or forced, is deterministically chaotic with an attractor as solution in the phase space. The transition from spontaneous to forced laser droplet generation is thus an example of an experimental realization of a chaos-to-chaos transition with an intermediate dynamically nonstationary phase. Notably, although the laser droplet generation is governed by additional physical phenomena, including light-metal interaction, heating and phase transitions, the dynamics of the process is similar to the one observed in traditional dripping faucet experiments. In addition, the presented results indicate that nonlinearity is an innate ingredient of laser droplet generation, which will be taken into account in future modeling and controlling attempts. We hope that acquired deeper understanding of the examined process will be of value when striving towards the integration of the process into outlined industrial applications.

\nonumsection{Acknowledgments} \noindent We acknowledge support from the Slovenian Research Agency (Program P2-0241; Grant Z1-2032) and the COST-P21 action Physics of Droplets.

\end{document}